# Report from KEK

High gradient study results from Nextef

Toshiyasu Higo[1*]

KEK, High Energy Accelerator Organization – Accelerator
1-1 Oho, Tsukuba, Ibaraki, 305-0801 – Japan

Most up-to-date high gradient test of the CLIC prototype structures as of September 2011 is described in this report. The "T24" undamped structure showed fast processing time, still-decreasing breakdown rate and its breakdown rate was estimated to be as low as the CLIC requirement. The "TD24" damped structure showed not so excellent high gradient performance as undamped "T24" but the characteristics was much improved than the damped "TD18" structure with higher magnetic field. Further R&D is needed and we present some of the present efforts at KEK.

## 1   Introduction

The high gradient studies based on X-band technology has been pursuing at KEK in close collaboration [1] with SLAC and CERN. One of the recent focus is a series of tests of CLIC prototype structures [2]. First test was on "T18" undamped structure. We established the experimental system with this. Then we tested "TD18" damped structure. We observed almost two order of magnitude higher breakdown rate[3]. We speculated that this high rate came from its parameter choice, especially high magnetic field. An improved design parameters were realized in 24-cell structure [4], "TD24." We firstly tested undamped version "T24" of this design[5]. This experiment was interrupted by the Earthquake in Tohoku area. After recovered from the damage due to the earthquake, we tested "T24" but it showed degradation of high gradient performance, probably due to the exposure to the air. Finally and up to now, we have been testing the damped "TD24" structure. In this paper, we focus on these two accelerator structure, "T24" and "TD24" and try to evaluate the basic performance with respect to the CLIC criteria on breakdown rate.

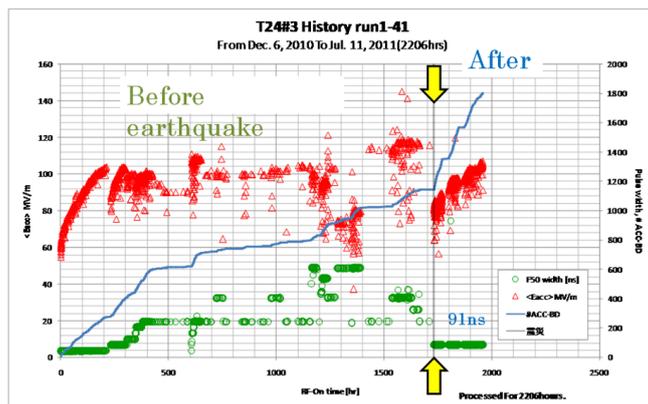

Figure 1: Whole processing history of "T24"

## 2   "T24" result

The whole processing and long-run is shown in the Figure 1. It was processed to 100MV/m level within 200 hours at the short pulse width. This is actually fastest of four structures, among undamped and damped, 18-cell and 24-cell structures, as shown in Figure 2. Here the accelerator field is shown as function of integrated number of accelerator-origin fault, called ACC-BD, accelerator breakdown.

After exposed to the air at the timing shown in yellow arrow in Figure 1, it showed very poor performance even at the operation with the short pulse width.

## 2.1 Evolution of breakdown rate

Figure 3 shows the breakdown rate (BDR) of "T24" as function of accelerator field. It was evaluated at several timings of the whole operation period. The nominal time elapsed before such evaluation counted from the beginning of processing is shown in the figure in hours. The dashed lines are for guiding eye and its slope was taken from the fitting of the first evaluation at 400 hours. If the slope does not change much, those data points in later timing indicate that it can run with meeting the CLIC criteria ($3 \times 10^{-7}$ bpp/m) at 100 MV/m. Note that the pulse width of the present operation for this test was 252 nsec, while CLIC operates at 156 nsec in its flat top with slow ramping at the beginning. This may even decrease the BDR of the actual pulse shape.

If we read those value which these dashed lines intersect to the 100 MV/m line, the result shows clearly the reduction of the BDR of this structure, as shown in Figure 4. It showed actually ever decreasing feature even later than 1500 hours of operation.

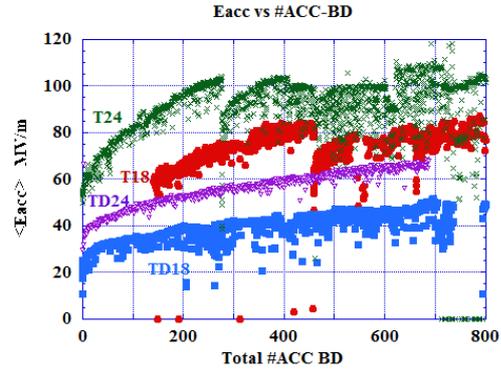

Figure 2: Comparison of initial processing

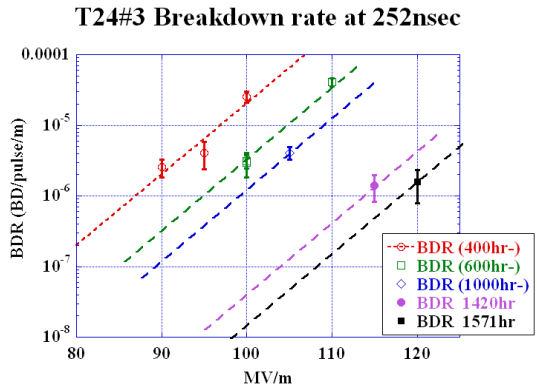

Figure 3: BDR of "T24" vs Eacc

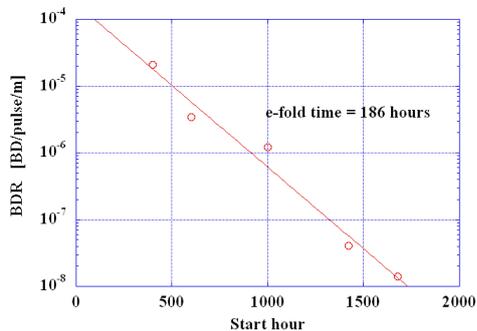

Figure 4: Evolution of BDR of "T24"

## 2.2 Deterioration due to arcing

The vacuum discharge may modify the copper surface of the accelerating cell. It will detune the cell frequency, resulting in a change of total phase advance through the structure. The total phase advance along the whole structure of T24 was measured in a low power setup. As shown in Figure 5, it showed the phase shift decrease of 2 degrees along the structure "T24," which is equivalent to the frequency increase of 0.16 MHz in average if we take into account various environmental parameters during the measurement.

This measurement is susceptible to many perturbations, such as temperature and humidity,

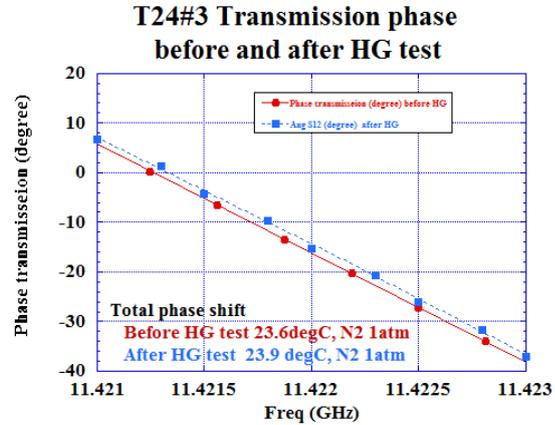

Figure 5: Total phase shift along structure.

and we admit that this measurement should be refrained to confirm the result in a condition with the smaller systematic errors. Note that his change should be small enough for the whole life of the accelerator structure.

## 3 Tests in simple geometry

A series of tests with simpler geometry is being prepared at KEK for studying the mechanism of breakdowns and for searching the suppression methods. The so-called "single-cell" setup [6] shown in Figure 6 is in mind. In this figure, RF power comes from right in a TM02 mode in a circular waveguide. The first cell, matching cell, adjusts the impedance from circular pipe to the three accelerator cells. The center cell has highest field. The last cell behaves just the same as the first, but stops the power to left keeping the accelerator geometry kept for the center cell. This figure shows the case of damped cell as the center cell

A small concrete vault is being prepared for this test as shown in Figure 7.

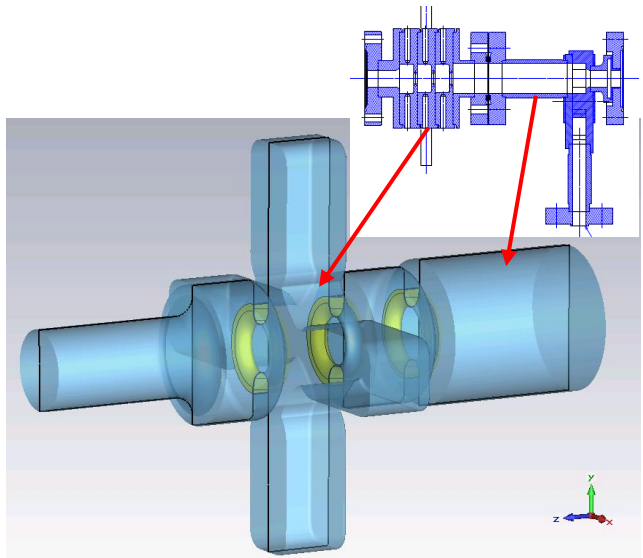

Figure 6: Test in simple geometry.

## 4 Acknowlegment

Most of the present study has been proceeding under a tight collaboration among CERN, SLAC and KEK. The authors greatly thank the DG's, heads of accelerator division and management people of these laboratories. We also give much credit to those who have been working with us in the above collaboration.

## 5 References

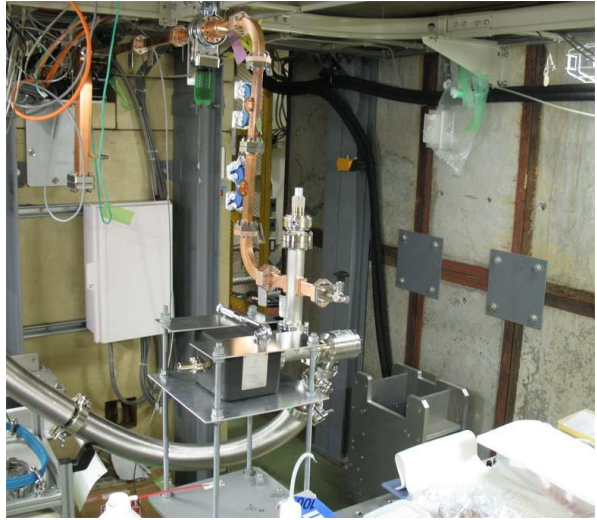

Figure 7: New concrete vault for test.